# Gravitational Gamma Spectrometer for Studying the Gamma Resonance of the Long-Lived Isomer [103m]Rh

A.V. Davydov, Yu.N. Isaev, V.D. Kalantarov, M.M. Korotkov,
V.V. Migachev, Yu.B. Novozhilov, A.M. Stepanov

State Scientific Center of Russian Federation Alikhanov Institut for Theoretical and Experimental Physics NRC "Kurchatov Institut"

1. **Introduction.**

After the successful experiments on the observation of the gamma resonance of the long-lived isomer [109m]Ag [1,2] we decided to study the gamma resonance of the more long-lived isomer [103m]Rh. Nuclide [103]Rh has an excited isomeric state with energy 39.75 keV and the half-life equal to 56.12 min [3], that about 84 times exceeds the half-life of the [109m]Ag isomer. The natural width of gamma lint emitted by rhodium isomer is less than that of silver isomer by the same 84 times. This difference does not permit to use for studying gamma resonance of rhodium isomer the gravitational gamma spectrometer which was used earlier in the experiments with silver isomer, and requires the fabrication of a new device with essentially higher sensuality for the angle of gamma beam inclination that is the higher degree of collimation of gamma beam. Rhodium metal has the Debye temperature equal to 480 K [4], that in combination with low energy of emitted gamma quanta would permit as it would seem to perform the Mössbauer experiments with rhodium isomer at room temperature because the probability of the recoilless gamma ray emission (absorption) is equal to 0.465 under these conditions. However there are theoretical works [5,6] in which it is stated that the broadening of Mössbauer gamma line is possible as result of Doppler effect of second order strongly arising with temperature. The experiments with isomer [109m]Ag were performed at liquid helium temperature and this effect could not be observed. But there were the other reason for gamma line broadening – dipole-dipolar interaction of nuclear magnetic moments, which would broaden the natural width of gamma line about $10^5$ times. However the experiments performed with gamma sources fabricated by the sparing technology of thermo-diffusion introducing of parent nuclide $^{109}$Cd into silver [1,2,7-10] shown that such broadening is absent. A.V. Davydov proposed [11] that its reason consists in the much more quicker changes of the energy of dipole-dipolar interaction by the value and by sign than the processes of gamma quantum emission and absorption by nuclei are running. If this proposition is correct then the broadening via Doppler effect of the second order must be absent by the same reason. If however the gamma line of rhodium isomer would turn out to be broadened then the we should be forced to perform the work with rhodium isomer at liquid helium temperature.

2. **Operation principle of gravitational gamma spectrometer**.

Given spectrometer does not differ by its operation principle from his prototype – gravitational gamma spectrometer with which the gamma resonance of $^{109m}$Ag was observed. It similarly registers the resonant gamma ray absorption in the substance of very gamma source.

The angular divergence of gamma beam in the experiments with silver isomer was determined by the sizes of gamma source and detector and by the distance between them. Registration of this divergence required to determine the slope angle for with respect to the horizontal plane of which gamma quantum way between gamma source and detector. Simultaneously the length of photon way was also determined in the source substance on which the resonance absorption of gamma rays could take place. One would do something like that in processing data of future experiments with given spectrometer. It will be described with more details in the following section.

If near the Earth surface the point 1 of gamma quantum emission and the point 2 of its possible resonant absorption by nucleus are separated in vertical direction by distance *H* then position of gamma resonance in the point 2 turns out to be shifted relatively the energy of gamma quantum emitted in the point 1 by the value

$$\Delta E = E_\gamma \frac{gH}{c^2} \quad (1)$$

In the absence of magnetic field the cross section of Mössbauer resonant absorption of gamma rays emitted in the horizontal direction is expressed by the formula

$$\sigma_0 = \pi \lambdabar^2 \frac{2J_e + 1}{2J_0 + 1} \cdot \frac{1}{1+\alpha_t} f^2 \quad (2)$$

Here $\lambdabar$ is the wave length of gamma radiation divided by $2\pi$, $J_e$ and $J_0$ are the spins of excited and ground nuclear states correspondingly, $\alpha_t$ – total coefficient of gamma ray inner conversion, $f$ – probability of recoilless gamma ray emission and absorption.

Cross section of resonant absorption of gamma rays with energy $E_\gamma$ under the condition that gamma line is broadened by $k$ times and has the real width equal to $\Gamma_{real} = \Gamma_{nat.} \cdot k$ where $\Gamma_{nat}$ is the natural width of gamma line, is expressed by the following formula

$$\sigma(E_\gamma) = \frac{\sigma_0}{k} \cdot \frac{\frac{\Gamma_{real.}^2}{4}}{(E_0 - E_\gamma)^2 + \frac{\Gamma_{real.}^2}{4}} \quad (3)$$

Here $E_0$ – energy corresponding to the maximum of gamma line intensity.

The normalized on 1 spectrum of source gamma rays shifted by energy *s* with respect to the absorption line has the view

$$N(E_\gamma)dE_\gamma = \frac{2}{\pi \Gamma_{real.}} \cdot \frac{\Gamma_{real.}^2 / 4}{(E_0 - E_\gamma - s)^2 + \frac{\Gamma_{real.}^2}{4}} dE_\gamma \quad (4)$$

Averaging of cross section (3) over the spectrum (4) gives

$$\sigma = \frac{\sigma_0}{2k} \cdot \frac{1}{1 + \frac{s^2}{\Gamma_{ecm.}^2 k^2}} \quad (5)$$

The values of $s$ and $\Gamma$ for $^{103m}$Rh isomer are the following:

$$S = 6.93 \cdot 10^{-26} H \text{ erg}, \quad \Gamma = 2.161 \cdot 10^{-31} \text{ erg}$$

Substituting these values in (5) we obtain

$$\sigma = \frac{\sigma_0}{2k} \cdot \frac{1}{1+1{,}028 \cdot 10^{11} \left(\frac{H}{k}\right)^2} \quad (6)$$

If the gamma beam to be registered is inclined as it is shown on fig. 1 then during the movement of photon along the inclined way in the source substance the vertical shift $H$ arises and this leads to the increase of s and to the corresponding decrease of Gamma ray resonant absorption probability.

If gamma quanta must run in the source substance the way equal to $l$ from its emission point in the plane of magnetic meridian in the direction inclined by the angle $\theta$ with respect of the horizontal plane then the intensity of these photons emitted by the source in the element of solid angle $d\varphi$ will be equal to

$$N_\gamma(d\varphi) = N(dV)\exp(-\mu_e l)\ \exp\left(-\frac{\sigma_0 \nu}{2k} \int_0^l \frac{dl}{1+1.028\cdot 10^{11}(\frac{l\sin\theta}{k})^2}\right) d\varphi \quad (7)$$

After the integration in the second exponent we have

$$N_\gamma(d\varphi) = N(dV)\exp(-\mu_e l)\exp\left(-\frac{\sigma_0 \nu}{2}\frac{1}{3.206\cdot 10^5 \sin\theta} arctg(\frac{3.206\cdot 10^5 \sin\theta}{k}L)\right)d\varphi \quad (8)$$

Here $N(dV)$ – the quantity of gamma quanta emitted in the source volume element $dV$, $\mu$ - the usual (electronic) coefficient of gamma ray absorption, $\nu$ – the number of atoms per 1 cm$^3$ which able to resonantly absorb gamma rays, $l$ is the length of the gamma quantum way in the source substance. If the nuclei of gamma source and of resonant absorber are in the identical magnetic fields splitting the emission and absorption gamma lines of isomer $^{103m}$Rh into non-overlapping components then the cross section (2) decrease by 64/17 times [12]. Just such these conditions are realized for isomer $^{103m}$Rh in the geomagnetic field for the horizontal gamma ray beam.

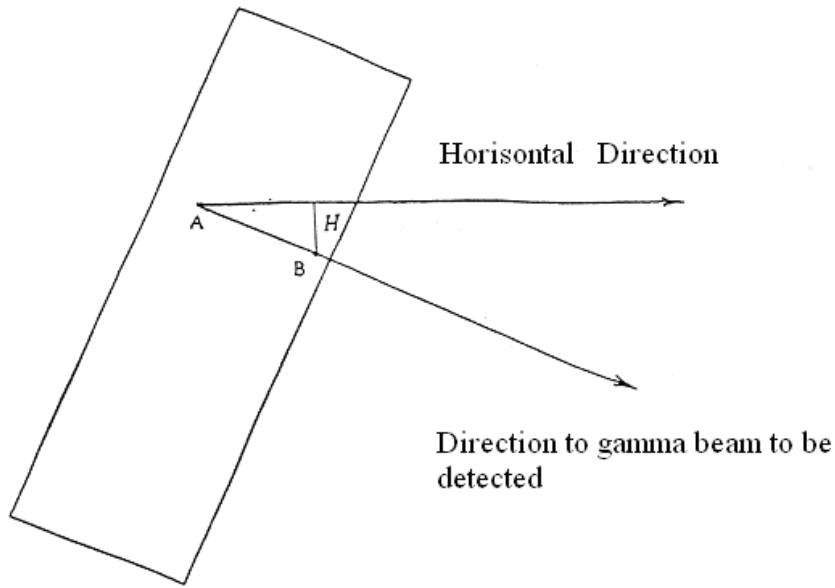

**Fig. 1**. Chart explaining the operation principle of gravitational gamma spectrometer. The cross section of gamma source is shown in the position when the gamma beam to be detected is inclined with respect of horizontal plane by the angle $\theta$.

A – point of gamma quantum emission, B – point of its possible resonant absorption, $H$ – the difference of vertical coordinates of these points increasing with the growth of the photon way in the gamma source substance.

3. **Construction of gravitational gamma spectrometer.**

Given spectrometer is a device of table-top type. Its general view is shown on the photography on fig.2. On the supporting plate 1 the plate 3 is fastened by means of hinge 2. Plate 3 may be inclined with respect of the horizontal plane by means of micrometric screw 4 supplied by the limb 5 with a scale of 360 degrees. On this moveable plate the holder of gamma source 6, brass pre-collimator 7 and main glass collimator 8 are attached. Scintillation detector 9 of type 8S8/2 made by firm ScintiTech (Massachusetts, USA) with NaI(Tl) crystal of 40 mm in diameter and 10 mm in thickness is also firmly fastened to this plate.

Collimating system of the spectrometer is described in our paper [13]. It consists of multi-slit brass pre-collimator (plates 240×80×1 $mm^3$ and slit aperture of 2 mm) and main glass multi-slit collimator made of the plates 60×60×0.2 $mm^3$ with slit aperture 0.2 mm. Spectrometer is so placed that its prolonged axis lies in the plane of magnetic meridian. The instrument is supplied by a system of Helmholz rings 10 which serves as an additional means for revealing of resonant gamma ray absorption. The frames of these rings are the duralumin bicycle rims 26 inches in diameter. The winding of each coil consists of 35 spires of wire PV-3 1 $mm^2$ in cross section. The coils are mounted on the aluminium constructive elements.

Spectrometer is so placed between the Helmholz rings that the gamma source is located in the center of ring system. One can with help of this system compensate the vertical component of terrestrial magnetic field at the place of location of gamma source. As it is shown in [12] the most favourable conditions arise for resonant gamma ray absorption when the beam is parallel or anti-parallel to the horizontal direction of magnetic field. On the fig. 3 the dependence is shown of the factor proportional to the probability of resonant gamma ray absorption on the angle between the direction of gamma quantum emission and the direction of magnetic field calculated in [12] acting on the nuclei emitting and resonantly absorbing gamma rays.

Knowing the natural direction of Earth magnetic field one can determine what relative part of total resonant absorption effect reveals in measurements of gamma ray intensities at the natural direction of Earth magnetic field and at the compensated vertical component of this field and find with its help the value of total resonant absorption effect.

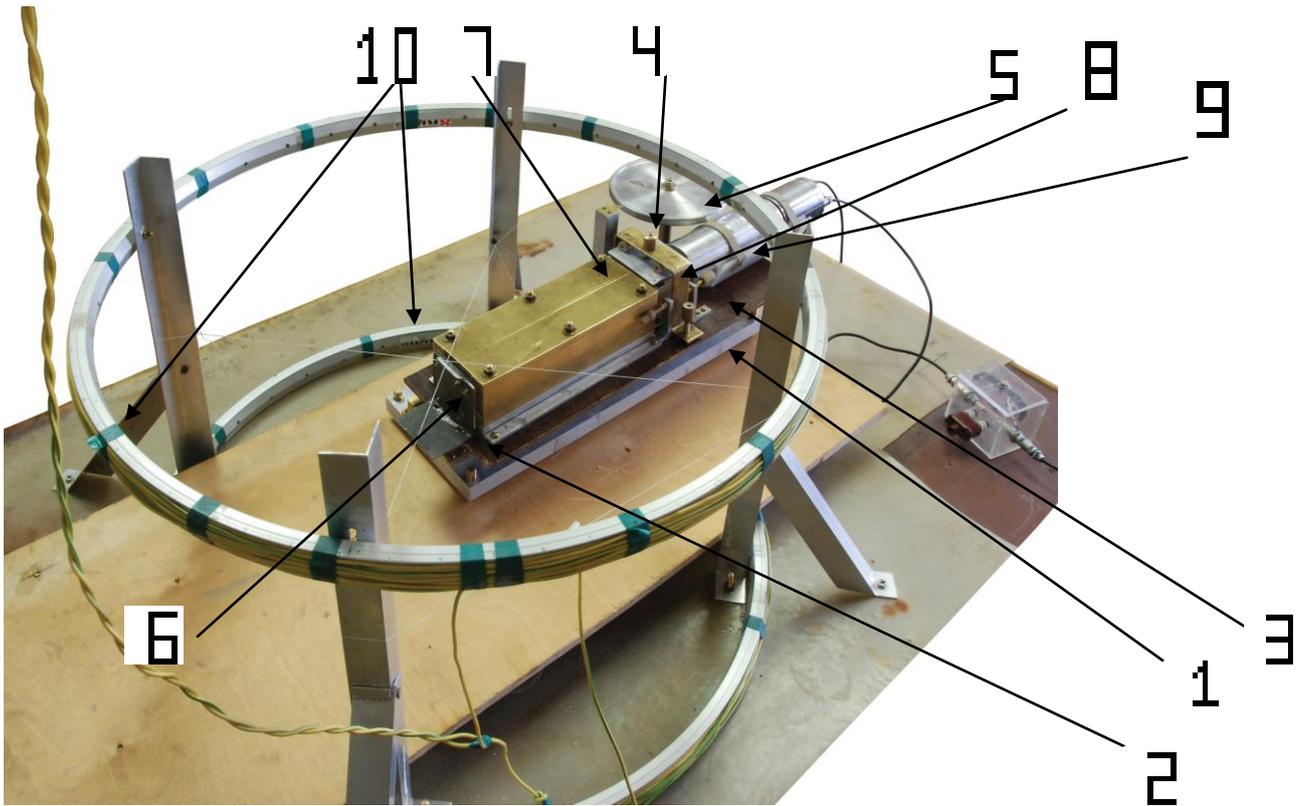

**Fig. 2**. General view of gravitational gamma spectrometer.
1 – supporting plate, 2 – hinge, 3 – rotating plate, 4 – micrometric screw, 5 - limb with a scale of 360 degrees, 6 - holder of gamma source, 7 - brass pre-collimator, 8 – main glass collimator, 9 - scintillation detector, 10 – Helmholz rings.

4. **Preparation of spectrometer for the operation**.

It follows from above written that work with given device must begin from additional magnetic measurements. It is required, firstly, determine the direction of the plane of magnetic meridian at the place of gamma source and direct the prolonged axis of spectrometer parallel to this plane. Secondly, determine the natural direction of Earth magnetic field acting on gamma source and at last find the current of Helmholz rings required for compensation of vertical component of terrestrial magnetic field at the place of gamma source. We were performing these measurements using the small size indicator of FIT-1 type designed and fabricated in VNIIFTRI. They are described in details in [14]. If the location of spectrometer would be changed these measurements must be repeated anew. The tuning of scintillation detector which resolving power depends on the applied voltage is the preliminary work also besides of magnetic measurements. Corresponding measurements performed with gamma rays of $^{109m}$Ag isomer with energy 88.03 keV have shown that the minimal value of measured gamma line width at half of its height turned out to be equal to ~ 8 keV at applied voltage of 600 volts, that is about 9 % of gamma ray energy.

5. **Problem of gamma source**.

There are several methods of gamma source fabrication for experiments with $^{103m}$Rh isomer. One of them consists in the irradiation of rhodium target by bremsstrahlung of electrons with energy 6-8 MeV. Just this method was used by Chinese group in their work [15]. It is clear however that this method is not convenient because the small life time of rhodium isomeric state. Even if the spectrometer is placed near the source of bremsstrahlung It would be necessary interrupt the measurements every 1-2 hours for a new irradiation of gamma source. At the transposition of target from distant place of irradiation the losses of gamma activity would be inevitable in addition to the multiple irradiations.

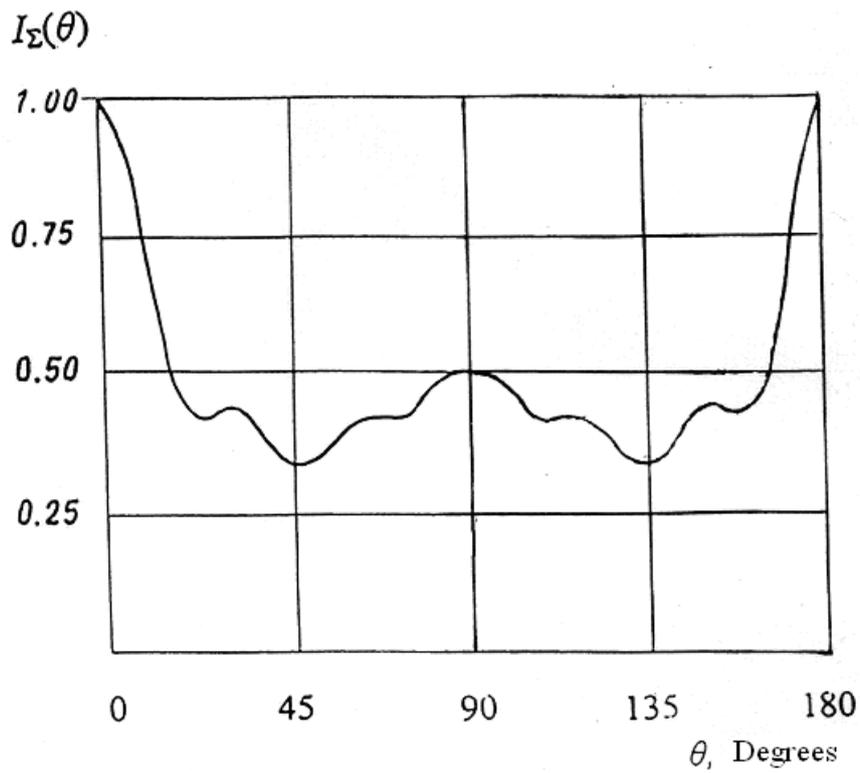

**Fig. 3**. Factor $I_\Sigma(\theta)$ which is proportional to the probability of gamma ray resonant absorption isomer $^{103m}$Rh in dependence on the angle $\theta$ between the direction of gamma ray emission and the direction the magnetic field

The second method consists in the fabrication of ruthenium-rhodium alloy and in irradiation it by thermal neutrons in the reactor. In this case the nuclear reaction $^{102}$Ru(n,γ)$^{103}$Ru will run. Nuclide $^{103}$Ru decays with half-life ~39 days on the excited isomeric state of $^{103}$Rh with energy 39.75 keV. The absorption of neutrons by other ruthenium isotopes does not lead to the appearance of essential background gamma activity.

The third method which is the most acceptable one costs in the irradiation of rhodium target by protons with energy sufficient for running of the reaction $^{103}$Rh(p,n)$^{103}$Pd. Nuclide $^{103}$Pd decays with half-life 16.96 days with the creation of $^{103}$Rh nucleus in the excited isomeric state with energy 39.75 keV. One may work with such gamma source during one month. The question on the optimal conditions for the irradiation the rhodium target by protons is considered in our paper [16]. On the fig. 4 the results are shown of measurements cross section of the reaction $^{103}$Rh(p,n)$^{103}$Pd in dependence on proton energy (EXFOR data). Mean value of the cross section in the macsimum this distribution is equal to 0.76 barn at proton energy 10.3 MeV. Mean width of the distribution presented on the fig. 4 on the half-height is equal to ~ 5.2 MeV. The ITEP liner accelerator gives the proton beam with energy 25 MeV and with mean current 3 – 5 mcA. If one uses the aluminium filter for decreasing the proton energy to 10.58 MeV and irradiate by this beam the rhodium target 0.1mm in thickness during three weeks the gamma sours may be obtained with intensity of ~ 4 $10^6$ photons/s. This give acceptable counting rate of scintillation detector which is placed at 30 cm from the sours and separated from it by collimating system.

Authors thank D-r V.A. Kalinin from Radium Institute of Russian Academy of Sciences for his help at the search for the information on the energy dependence of the cross section of $^{103}$Rh(p,n)$^{103}$Pd reaction.

This work was supported by RFBR (grant 13-02-01174a).

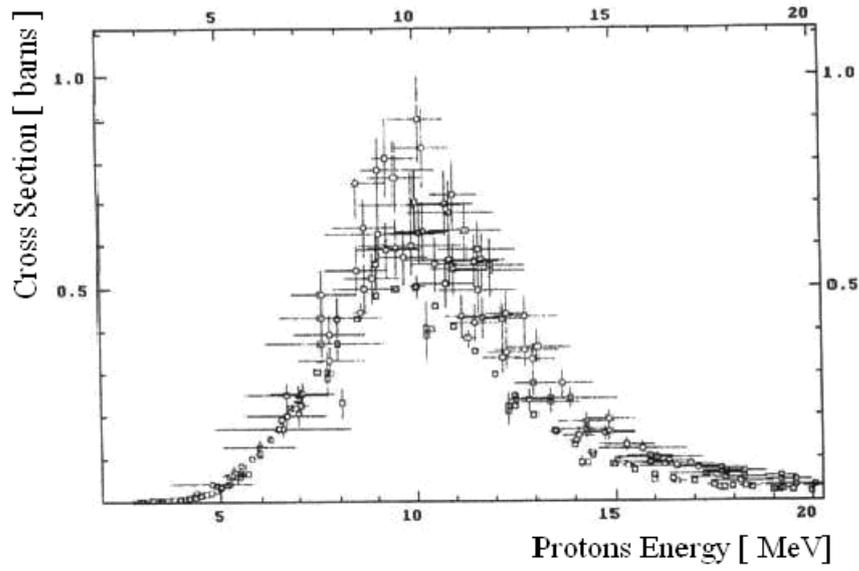

**Fig. 4**. The cross section of the reaction $^{103}Rh(p,n)^{103}Pd$ in Dependence on the proton energy.

Abstract

The principle of operation and the construction are described of the gravitational gamma spectrometer to study the gamma resonance of the long-lived isomer 103mRh. This is a table-top device which measures the form of gamma resonance by the dependence of counting rate of gamma-rays in collimated beams on the angle of their inclination with respect of the horizontal plane. Methods of fabrication of gamma-sources for this device are considered.

Captions to the figures.

Fig. 1. Chart explaining the operation principle of gravitational gamma spectrometer. The cross section of gamma source is shown in the position when the gamma beam to be detected is inclined with respect of horizontal plane by the angle $\theta$.

A – point of gamma quantum emission, B – point of its possible resonant absorption, $H$ – the difference of vertical coordinates of these points increasing with the growth of the photon way in the gamma source substance.

Fig. 2. General view of gravitational gamma spectrometer.
1 – supporting plate, 2 – hinge, 3 – rotating plate, 4 – micrometric screw, 5 - limb with a scale of 360 degrees, 6 - holder of gamma source, 7 - brass pre-collimator, 8 – main glass collimator, 9 - scintillation detector, 10 – Helmholz rings.

Fig. 3. Factor $I_\Sigma(\theta)$ which is proportional to the probability of gamma ray resonant absorption isomer $^{103m}$Rh in dependence on the angle $\theta$ between the direction of gamma ray emission and the direction the magnetic field.

Fig. 4. The cross section of the reaction $^{103}$Rh(p,n)$^{103}$Pd in Dependence on the proton energy.

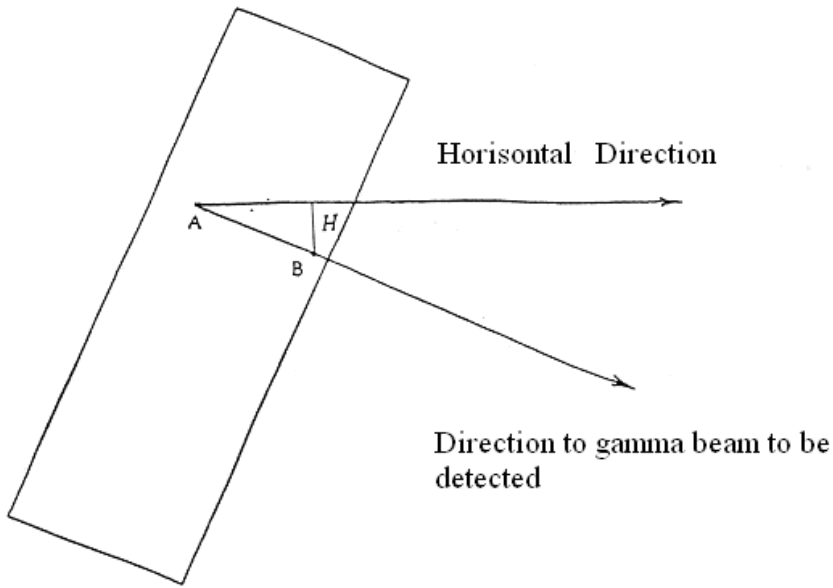

**Fig. 1**. Chart explaining the operation principle of gravitational gamma spectrometer. The cross section of gamma source is shown in the position when the gamma beam to be detected is inclined with respect of horizontal plane by the angle $\theta$.

A – point of gamma quantum emission, B – point of its possible resonant absorption, $H$ – the difference of vertical coordinates of these points increasing with the growth of the photon way in the gamma source substance.

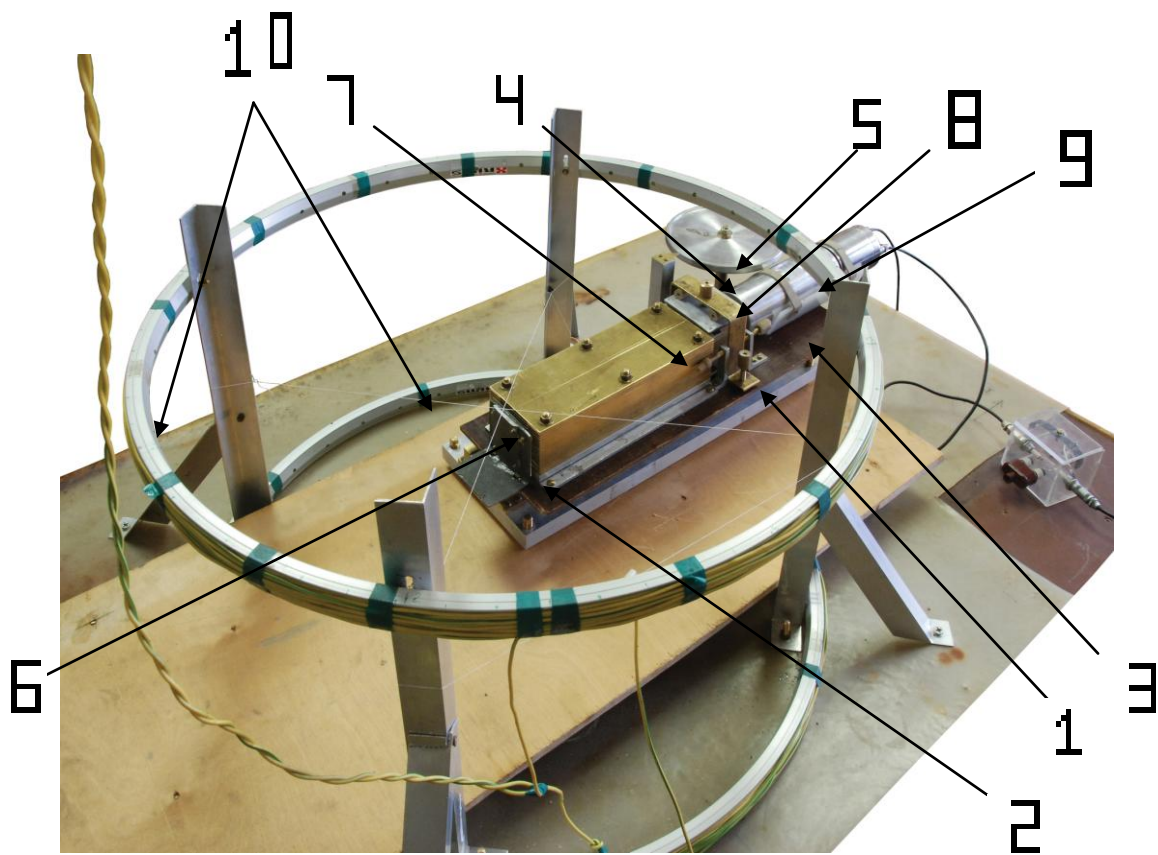

**Fig. 2**. General view of gravitational gamma spectrometer.
1 – supporting plate, 2 – hinge, 3 – rotating plate, 4 – micrometric screw, 5 - limb with a scale of 360 degrees, 6 - holder of gamma source, 7 - brass pre-collimator, 8 – main glass collimator, 9 - scintillation detector, 10 – Helmholz rings.

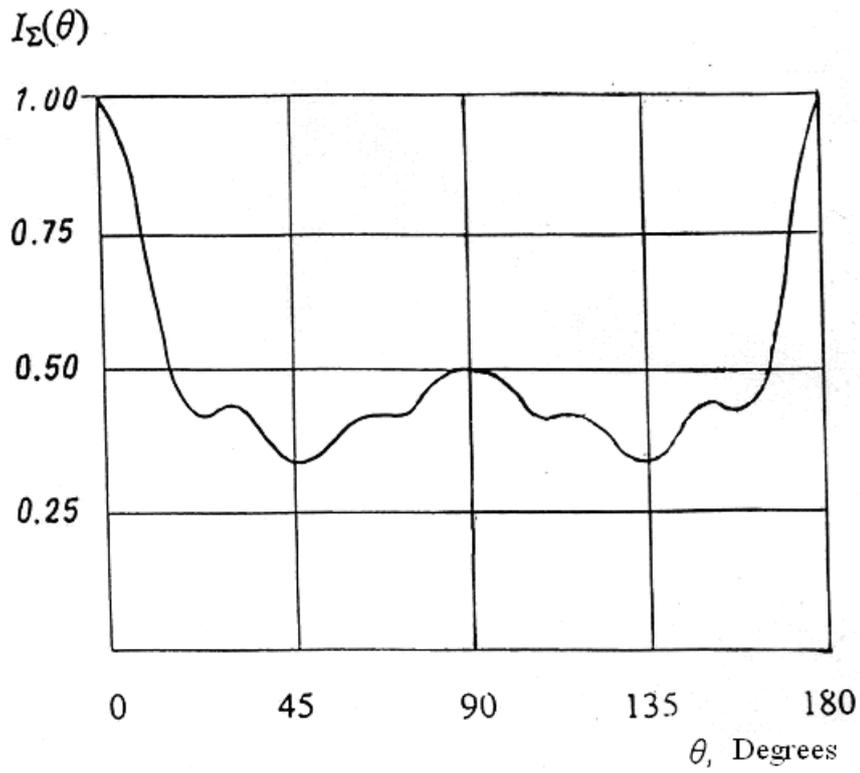

**Fig. 3**. Factor $I_\Sigma(\theta)$ which is proportional to the probability of gamma ray resonant absorption isomer $^{103m}$Rh in dependence on the angle $\theta$ between the direction of gamma ray emission and the direction the magnetic field

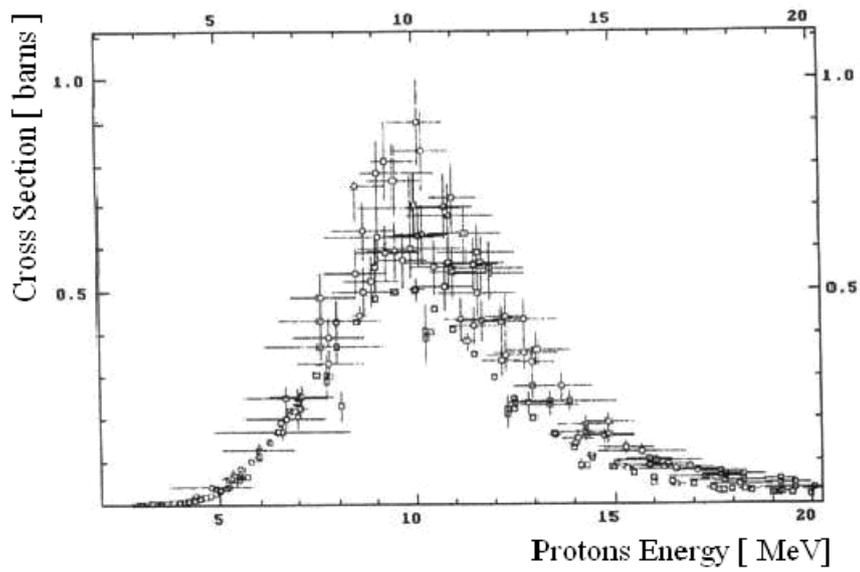

**Fig. 4**. The cross section of the reaction $^{103}$Rh(p,n)$^{103}$Pd in Dependence on the proton energy.